\pdfoutput=1

\documentclass[11pt]{article}

\usepackage[final]{acl}

\usepackage{times}
\usepackage{latexsym}

\usepackage[T1]{fontenc}

\usepackage[utf8]{inputenc}

\usepackage{microtype}

\usepackage{inconsolata}

\title{Blind Spots and Biases: Exploring the Role of Annotator Cognitive Biases in NLP}

\author{Sanjana Gautam\\
    Information Sciences and Technology \\
    Pennsylvania State University, USA\\
  \texttt{sanjana.gautam@psu.edu}\\
    \And
    Mukund Srinath \\
    Information Sciences and Technology \\
    Pennsylvania State University, USA\\
  \texttt{mukund@psu.edu}}

\begin{document}
\maketitle
\begin{abstract}
With the rapid proliferation of artificial intelligence, there is growing concern over its potential to exacerbate existing biases and societal disparities and introduce novel ones. This issue has prompted widespread attention from academia, policymakers, industry, and civil society. While evidence suggests that integrating human perspectives can mitigate bias-related issues in AI systems, it also introduces challenges associated with cognitive biases inherent in human decision-making. Our research focuses on reviewing existing methodologies and ongoing investigations aimed at understanding annotation attributes that contribute to bias.
\end{abstract}

\section{Introduction}

With the recent rapid expansion of generative AI models, we have witnessed their numerous benefits and the emergence of substantial ethical concerns \cite{thoppilan2022lamda, rudolph2023chatgpt}. There has been an influx of remarkable and noteworthy work that describes the issues of fairness, toxicity and bias in the text generation process \cite{bender2021dangers, abid2021persistent, seaborn2023m}. Several models are deployed as real-world solutions with a lack of informed consideration of their social implications, especially in sensitive fields such as healthcare, journalism, law, and finance \cite{khowaja2023chatgpt}. Recent research has revealed that these language models can mimic human biases present in language, perpetuating prejudiced behaviour that dehumanizes certain socio-demographic groups by deeming them more negative or toxic \cite{havens2022uncertainty, blodgett2020language}. 

One of the proposed solutions to this issue has been to introduce human annotators to label the training corpora or validate pre-labelled datasets and manually remove toxic (or biased) data entries \cite{havens2022uncertainty, cabrera2014systematic}. It is common practice for machine learning systems to rely on crowd-sourced label data for training and evaluation \cite{wu2022survey}. It is also well-known that biases present in the label data can induce biases in the trained models \cite{hettiachchi2021investigating}. Therefore, while humans-in-the-loop for model training may seem like an intuitive solution, it often introduces additional biases due to inherent cognitive biases in humans \citep{parmar2022don}. Crowdwork annotation studies conducted on MTurk (and other crowdwork platforms) where the participants come from a specific demographic population can potentially perpetuate populist viewpoints \citep{reinecke2015labinthewild}. 

Prior work has established the pitfalls in human rationality, as influenced by the lived experiences and environment, which Herbert Simon termed bounded rationality \cite{simon1957models}. Human biases have been identified to be the resulting gap between rational behaviour and heuristically determined behaviour \cite{tversky1974judgment, bojke2021developing}. Over 180 cognitive biases have been identified, spawning everything from social interaction to judgment and decision-making with research spanning over 70 years \cite{talboy2022reference}. These tendencies or patterns can lead to faulty reasoning, irrationality, and potentially detrimental outcomes.

Bias sometimes emerges due to distractions, lack of interest, or laziness among annotators regarding the annotation task, leading them to select inaccurate labels. However, more concerning is the label bias stemming from informed and well-intentioned annotators who consistently exhibit disagreement \cite{hovy2021five}. \citet{plank2014linguistically} demonstrated that this form of bias emerges when multiple correct labels are possible. For instance, the term 'social media' can be legitimately interpreted either as a noun phrase consisting of an adjective and a noun or as a noun compound comprising two nouns. For example, \citet{sap2019risk} demonstrated that these biases mirror social and demographic variances. For instance, annotators tend to evaluate utterances from various ethnic groups disparately and may misinterpret harmless banter as hate speech due to their unfamiliarity with the communication norms of the original speakers.

Merely relying on a few gold-standard corpora as training datasets or debiasing datasets is not a sustainable long-term strategy since languages undergo constant evolution. Thus, even a comprehensive sample can only encapsulate a momentary snapshot, offering at best a transient solution \cite{fromreide2014crowdsourcing}. We believe the design and set-up of the crowd work task plays a pivotal role in determining the goodness of data. In this work, we look at bias-diminishing strategies and identify the pressing questions in this area. Our central goal is to show that there is a need for standardized design principles when it comes to designing crowdwork studies. Specifically, we concentrate on the need for an HCI perspective in natural language processing research.

\section{Bias in AI Models}

Generative AI's propensity to amplify existing biases and create new ones has attracted considerable attention across a range of communities, including academics, policy-makers, industry, and civil society. Much of the initial work focused on developing quantitative definitions of fairness  \cite{dwork2012fairness, hardt2016equality, joseph2016fairness, liu2017calibrated, verma2018fairness}, and various technical methods for ‘debiasing’ AI models \cite{agarwal2018reductions, bolukbasi2016man, friedler2014certifying, zafar2017fairness}. When referring to de-biasing, we use the definition \textit{'removing undesired skews in the data and the model outcome, such as by equalising a metric of interest between groups'}. \textit{“Unintended bias”} is used to describe the different sources of bias that are introduced throughout an AI development life cycle \cite{lee2021algorithmic, suresh2021framework}, focusing on not just the bias introduced, but also the harm it causes \cite{crawford2016artificial}.

Recent studies have shifted focus from merely identifying sources of bias in AI, such as flawed data collection methods, to exploring the various harms caused by these biases. This shift is supported by interdisciplinary research that highlights the contextual nature of fairness. Factors such as regional and cultural differences in lived experiences significantly influence perceptions of fairness, revealing that certain algorithmic behaviours may only be deemed harmful in specific social or cultural contexts \cite{green2018myth, lee2021landscape, sambasivan2021re, selbst2019fairness}. Given these complexities, it is broadly acknowledged that eliminating bias or ensuring absolute fairness in AI systems is unfeasible \cite{kleinberg2016inherent, mehrabi2021survey, pleiss2017fairness}. Instead, the objective is to minimize fairness-related harms and other adverse impacts to the greatest extent possible \cite{mehrabi2021survey, selbst2019fairness, sun2019mitigating}. This perspective is further enhanced by recent interdisciplinary studies \cite{lewicki2023out}, which underscore the nuanced and multifaceted nature of fairness in AI.

Identifying and acknowledging systemic biases in data collection is a crucial step in mitigating their impact on the systems that are trained using this data, and is a critical prerequisite for achieving fairness in algorithmic decision-making \cite{hajian2016algorithmic}. While humans are integral to the system, participating in data collection and various phases thereof, it is imperative to emphasize that human computation \cite{quinn2011human}, the practice of harnessing human intelligence and cognitive abilities as computational elements, holds potential for addressing and mitigating these challenges.

\section{Cognitive Bias among Annotators}


As emphasized by \citet{van2023chatgpt}, ensuring human accountability is essential in scientific practice. The history of Large Language Models (LLMs) has shown that they can produce inaccurate information, or "hallucinations." To guarantee the accuracy of information, it is necessary to implement a rigorous verification and fact-checking process led by experts. Consequently, the discourse highlights the critical need for accountability in human-in-the-loop systems, particularly in response to the new challenges posed by these systems.

The importance of understanding and mitigating biases in crowd data is highly relevant to researchers, and others who rely on crowd data for creating automated systems. Prior work has explored various approaches to promoting fairness in machine learning, including the direct utilization of crowdsourced data \cite{balayn2018characterising}, leveraging crowds to assess perceived fairness of features \cite{van2019crowdsourcing, van2021effect}, applying pre-processing techniques such as removing sensitive attributes, resampling data to remove discrimination, and iteratively adjusting training weights for sensitive groups \cite{calmon2017optimized, kamiran2012data, krasanakis2018adaptive}, as well as employing active learning methods \cite{anahideh2022fair}.

The use of crowdsourcing for tasks such as data annotation can inadvertently introduce cognitive biases, stemming from the inherent design of the task itself. We have identified three primary reasons why annotated data can be problematic: (1) Unethical spammers submit imprecise or even arbitrary labels in order to maximize their financial advantage \cite{eickhoff2012quality} or due to external distractions. (2) Unqualified workers are, despite their best efforts, unable to produce an acceptable annotation quality \cite{eickhoff2014crowd}. (3) Malicious workers purposefully aim to undermine or influence the labelling effort \cite{wang2013characterizing}. However, we propose that there might be some factors that have not been uncovered in prior literature. Crowd-workers have their tasks cut out for them, in cases where the nature of task design causes the propagation of bias. Research on crowd work has often focused on task accuracy whereas other factors such as biases in data have received limited attention \cite{hettiachchi2021investigating}. 

Cognitive biases originate from individuals' own \textit{"subjective social reality"} which is often a product of lived experiences. This makes cognitive bias a deviation from the rationality of judgement, therefore it may consist of perceptions of other people that are often illogical \cite{martie2005evolution}.  An individual’s construction of social reality, instead of the objective input, may dictate their behaviour and lead to perceptual distortion, inaccurate judgment, illogical interpretation, or irrationality \cite{bless2014social}. Past work has demonstrated that cognitive bias can affect crowdsourced labour and lead to significantly reduced result quality. This performance detriment is subsequently propagated into system ranking robustness and machine-learned ranker efficacy  \cite{eickhoff2018cognitive}.


The annotation instructions provided to crowdworkers can inadvertently prime them to exhibit biases towards or against specific domain information, which can be exacerbated by poorly designed instructions. Furthermore, annotators are often not fully informed about the true purpose of the research, leading to an ambiguity effect that can make the decision-making process appear more challenging and less appealing due to the limited information available \cite{ellsberg1961risk}. Additionally, the phased revelation of information to annotators can result in an anchoring effect, where certain pieces of information are given disproportionate attention based on the timing of their disclosure. This underscores the importance of designing annotation studies that mitigate cognitive biases among workers, ensuring that the annotation process is fair, transparent, and unbiased.

\section{Crowd Control}

Humans in the loop bring a lot of value to generative AI and AI systems. Therefore, the solution to the issue of cognitive bias cannot be to remove the annotators from the system. Human annotators often bring expert judgements, that are valuable in creating ground truth labels. For example, annotation of medical imagery cannot be performed without the help of annotators who are medical professionals. Expert guidance, lived experiences and proximity to the problem domain make human annotators irreplaceable in the AI-training life-cycle. The common strategies of accounting for biases of annotators by employing qualification tests, demographic filters, incentives, and sophisticated worker models may not be enough to overcome this source of noise. There is therefore a need to control the annotation task design settings, to minimize the introduction of biases due to the cognitive biases of annotators. 
While cognitive biases and their effects on decision-making are well-known and widely studied, we note that AI-assisted decision-making presents a new decision-making paradigm. It is important to study their role in this new paradigm, both analytically and empirically.

Crowdwork platforms are often designed to position crowdworkers as interchangeable \cite{irani2013turkopticon}. While some forms of digital work can be decomposed and distributed, the presumption that all crowdsourced dataset annotators exercise near-identical capacities of perception and judgement ignores the fact that social position, identity, and experience shape how annotators' actions.

Previous research has highlighted the significance of the annotator population and the power dynamics inherent in platform-mediated crowdwork, both of which can perpetuate cognitive biases \cite{diaz2022crowdworksheets}. Building upon this foundation, we propose a novel framework to enhance transparency and robustness in the process of designing a crowdwork task. This approach holds promise for mitigating the impact of cognitive biases in crowdwork, thereby contributing to more reliable and trustworthy outcomes.

\section{Counter-measures for Biases}



To minimize bias in NLP annotation tasks, several steps can be implemented. Firstly, recruiting a diverse group of annotators from various backgrounds can help balance individual biases. Providing clear and detailed guidelines ensures uniform understanding across annotators. Training sessions, followed by calibration discussions, align annotator interpretations and reveal guideline ambiguities. An iterative feedback loop allows for regular quality checks and guideline adjustments based on annotator experiences. Measuring inter-annotator agreement with metrics like Cohen's Kappa highlights discrepancies and areas needing clarification. Annotation tasks should be designed to minimize bias, such as by rotating text assignments among annotators to avoid topical biases. Finally, a post-annotation analysis can detect any remaining biases, ensuring the reliability and fairness of the annotated data. 

However, biases can arise at any point in the AI lifecycle. It is therefore imperative for researchers to maintain a meticulous approach throughout the entire research process, encompassing various facets such as the selection of appropriate datasets, adherence to annotation schemes or labelling procedures, thoughtful considerations regarding data representation methodologies, judicious selection of algorithms tailored to the task at hand, and rigorous evaluation protocols for automated systems. Furthermore, researchers must consider the tangible real-world applications of their research endeavors. Particularly noteworthy is the imperative to consciously direct efforts towards leveraging technological advancements to uplift and empower marginalized communities, as underscored by \citet{asad2019academic}. Several studies critique existing bias mitigation algorithms for their lack of effectiveness due to inconsistent study protocols, inappropriate datasets, and over-tuning to specific test sets. To overcome these limitations, research needs to introduce robust evaluation protocol, and sensible metrics designed to evaluate algorithm robustness against various biases \citep{shrestha2022investigation}. 

Our future work derives from the insights presented in the preceding discussion. It posits that the roots of bias within AI systems often traced back to the initial stages of the annotation process, particularly during the instruction phase. Although not all cognitive biases are inherently detrimental, a pressing need exists to advance our comprehension of how to devise annotation studies that align with the principles of human-computer interaction (HCI). 

Our objective in this research endeavour is to contribute substantively to the ongoing efforts aimed at mitigating bias in crowd work. We intend to achieve this by focusing on the refinement of study design and instructional strategies. By incorporating insights from the HCI discipline, we aim to cultivate a nuanced understanding of how to create balanced annotation studies that minimize the emergence of bias. Through this work, we aspire to not only shed light on the pivotal role played by the annotation phase in propagating or mitigating bias but also to provide practical recommendations and guidelines for researchers and practitioners engaged in AI development and crowd work.

\section{Conclusion}

Our research highlights the critical importance of considering annotation attributes that contribute to bias in AI systems. The cognitive biases of annotators, inherent in human decision-making, can perpetuate and even amplify existing social disparities in AI models. To mitigate these issues, a multidisciplinary approach is necessary not only in deploying AI models but also in designing better systems for annotation tasks. By bringing together experts from diverse fields, including human-centered design, ethics, social sciences, law, healthcare, AI/ML, education, communication, and community representation, we can design annotation systems that are more inclusive, transparent, and fair. This collaborative framework is essential for developing annotation tasks that are free from biases, ambiguous, and unclear instructions, and that take into account the complexities of real-world data. Furthermore, a multidisciplinary approach is crucial for deploying AI models that are developed using these annotated data, ensuring that they are fair, transparent, and accountable. By acknowledging the limitations of human annotators and addressing them through a multidisciplinary approach, we can work towards a more equitable digital landscape where AI systems benefit both individuals and society as a whole.

\bibliography{custom}

\appendix

\end{document}